\documentclass[journal]{IEEEtran}

\usepackage{balance}       
\usepackage{graphicx}      
\usepackage{amsmath}            
\usepackage[pdflang={en-US},pdftex]{hyperref}
\usepackage{color}
\usepackage{booktabs}
\usepackage{textcomp}
\usepackage{multirow}

\ifCLASSINFOpdf
\else
\fi

\begin{document}

\title{VoCoG: An Intelligent, Non-Intrusive Assistant for Voice-based Collaborative Group-Viewing}

\author{\IEEEauthorblockN{Sumit Shekhar\IEEEauthorrefmark{1},
Aditya Siddhant\IEEEauthorrefmark{2},
Anindya Shankar Bhandari\IEEEauthorrefmark{3}, 
Nishant Yadav\IEEEauthorrefmark{4}}\\
\IEEEauthorblockA{\IEEEauthorrefmark{1}Adobe Research}
\IEEEauthorblockA{\IEEEauthorrefmark{2}Carnegie Mellon University}
\IEEEauthorblockA{\IEEEauthorrefmark{3}Columbia University}
\IEEEauthorblockA{\IEEEauthorrefmark{4}University of Massachusetts, Amherst}
\thanks{Manuscript received June 10th, 2018.
Corresponding author: S. Shekhar (email: sushekha@adobe.com).}}

\markboth{ArXiV}%
{Shekhar \MakeLowercase{\textit{et al.}}}

\IEEEtitleabstractindextext{%
\begin{abstract}
There have been significant innovations in media technologies in the recent years. While these developments have improved experiences for individual users, design of multi-user interfaces still remains a challenge. A relatively unexplored area in this context, is enabling multiple users to enjoy shared viewing (\textit{e.g.} deciding on movies to watch together). In particular, the challenge is to design an intelligent system which would enable viewers to explore together shows or movies they like, seamlessly. This is a complex design problem, as it requires the system to (\textit{i}) assess affinities of individual users (movies or genres), (\textit{ii}) combine individual preferences taking into account user-user interactions, and (\textit{iii}) be non-intrusive simultaneously. The proposed system VoCoG, is an end-to-end intelligent system for collaborative viewing. VoCoG incorporates an online recommendation algorithm, efficient methods for analyzing natural conversation and a graph-based method to fuse preferences of multiple users. It takes user conversation as input, making it non-intrusive. A usability survey of the system indicates that the system provides a good experience to the users as well as relevant recommendations. Further analysis of the usage data reveals insights about the nature of conversation during the interaction sessions, final consensus among the users as well as ratings of varied user groups.
\end{abstract}

}

\maketitle
\IEEEdisplaynontitleabstractindextext

\section{Introduction} \label{sec:Introduction}
The realm of human-computer interaction has vastly expanded with the technologies for immersive experience making great strides \cite{blascovich2011infinite}. Moreover, there has been a huge shift in the media consumption, with a large population shifting online for personalized consumption of media content, like video or music. Hence, there is a growing need for innovation in design of human-computer interaction techniques to provide a seamless immersive experience for media consumption \cite{Cesar2009FTIH, nathan2008collaboratv}.

A challenging design problem in this context is social/collaborative viewing, that aims to allow remotely located users to enjoy shared viewing of media content in a way that they feel being seated together, like conventional group viewing. The impact of group viewing on improving viewing experience has been well studied in television research \cite{lull1980social, webster1982impact}. The work by \cite{ducheneaut2008social} and \cite{nathan2008collaboratv} formalized the concept of remote social viewing. \cite{ducheneaut2008social} designed a  SocialTV experiment to investigate how groups behave when watching a program together. \cite{nathan2008collaboratv} built CollaboraTV, which incorporated user collaboration while watching television through messaging and shared interest profiles. In a large scale study of online sports viewing experience, Mo \textit{et al.} \cite{Ko2016TOCHI} demonstrated the effectiveness of sharing thoughts and information, and desire to be belonging to a group for improving the watching experience. Further, McGill \textit{et al.} \cite{McGill2016TOCHI} built a synchronous shared at-a-distance smart TV system, and analyzed the adoption of the system and the nature of communication. They also built a prototype in VR for shared viewing and showed its effectiveness in enhancing the viewing experience. Commercially, \textit{rabb.it} and \textit{togethertube.com} support synchronized viewing of broadcasted content. Otherwise too, most of the online video platforms support some form of social interaction. For example, Facebook Live allows user to "like" a live video, whereas Hulu enables users to edit and share video clips with other. The social functionality also helps users in content discovery on the platforms.
   
However, the design of an interface, which could meaningfully enable remote viewers to explore and decide video content they would like to watch together, has not been looked into extensively. While previous work enable remote users in a collaborative viewing session to communicate through chat, voice or video, there has been little focus on developing interfaces which would enhance content discovery experience in such scenarios. To this end, we present VoCoG, an intelligent system for  voice-based collaborative group-viewing. The proposed system attempts to address various challenges in achieving a seamless content discovery experience in collaborative viewing settings. 

Firstly, VoCoG incorporates voice as a medium of interaction between users. This is non-intrusive as the users are not required to type or click, and is particularly suited for immersive interfaces \cite{steuer1992define, sanchez2005presence}. Further, natural user conversations allow VoCoG to extract rich user feedback (like movie, star affinity, expressed sentiments, etc.) using advanced natural language processing techniques. Moreover, even though, popular personal assistants like Siri or Alexa are built out of voice-based interfaces, we believe that there has been limited work in voice-driven feedback-based recommendations in multi-user interaction systems. 

VoCoG deploys an online recommendation algorithm, which could efficiently update user preferences based upon the complex feedback from conversations. Conversation \cite{Converse2016KDD} and critique-based \cite{chen2012critique} and online \cite{bresler2014latent, zhao2013interactive, kawale2015efficient} recommendation methods have started gaining attention recently. We exploit insights from the recent work to build an online recommendation system, which computes the recommended movies for each individual, based upon the feedback from his/her conversations. 

Finally, the challenge is to how to combine the recommendations for each individual into the final watch list for the group as a whole. VoCoG uses the concepts for group behavior modeling in social network \cite{sherchan2013survey}, as well as for group-based recommendations \cite{masthoff2011group,mccarthy2006group} and takes into account for user-user agreements/disagreements, individual affinities towards movies, shows or stars as well as user behavioral traits, to arrive at meaningful recommendations for the group to watch. VoCoG can also detect if the group has reached a consensus on watching a video or not. 

The paper is divided into six sections. Section \ref{sec:RelatedWork} describes the related work in the area. The details of the design of the proposed interface, VoCoG are in Section \ref{sec:ProposedApproach}, while Section \ref{sec:SysUserInterface} describes the final prototype. A comprehensive user evaluation of the system is discussed in Section \ref{sec:SysEvaluation} followed by the conclusions in Section \ref{sec:Conclusions}.

\begin{figure*}
	\centering
\includegraphics[scale=0.3]{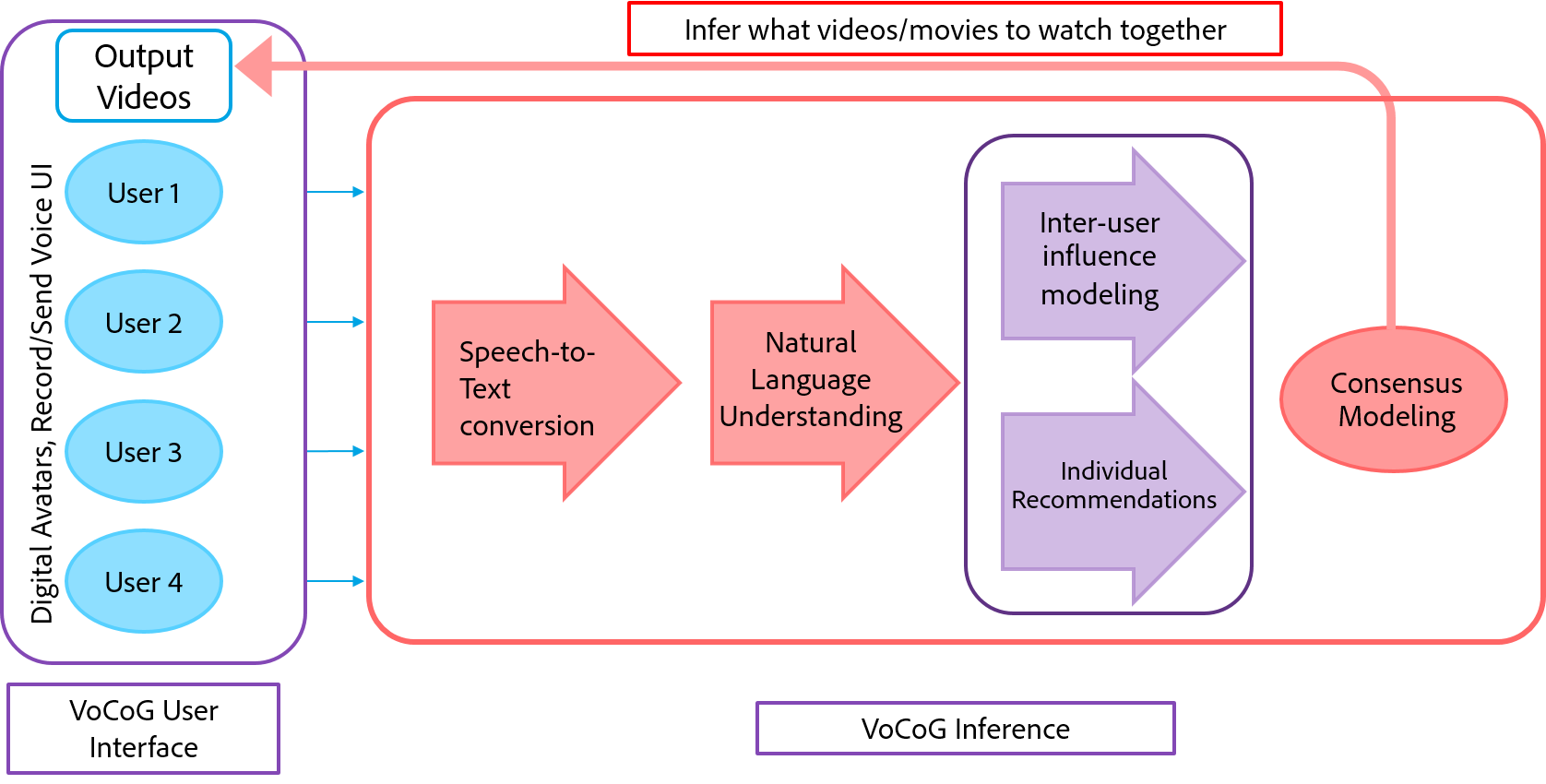}
\caption{Workflow diagram for VoCoG, an intelligent interface for collaborative group-viewing.}
\label{fig:VoCoGWorkflow}
\end{figure*}

\section{Related Work} \label{sec:RelatedWork}
In this section, we describe the related work in the design of multi-user interface, recommendations, conversation analysis and multi-user interaction modeling. \\
\textbf{Multi-user Interfaces:} There has been extensive research in the design of multi-user interfaces \cite{carlsson1993dive, curtis1994muds}. Virtual presence \cite{sanchez2005presence} as well as design of virtual world using avatars \cite{blascovich2011infinite} have been studied. The use of voice for human-machine interaction \cite{igarashi2001voice, cohen1995role} as well as immersive media \cite{salzman1999model} has been studied. There has been also related work in the domain of design of interactive shared viewing experience \cite{ducheneaut2008social, nathan2008collaboratv, becker2016interactive}. However, these approaches provide restrictive interaction mechanism between through chats or avatars. The proposed approach however is designed to account for rich conversation between users, and provide a seamless non-intrusive experience to the users.\\
\textbf{Recommendations:} A comprehensive survey of recommendation algorithms has been done by \cite{adomavicius2005toward}. There have been work for group-based recommendations \cite{roy2014exploiting,  masthoff2011group, mccarthy2006group}. Conversation-based \cite{Converse2016KDD, Wu2008Recsys} as critique-based  recommendations \cite{chen2012critique} have been studied. Online recommendations techniques like bayesian \cite{stern2009matchbox}, bandits \cite{zhao2013interactive}, latent analysis \cite{bresler2014latent} have been proposed recently. Our approach is motivated by \cite{bresler2014latent} to use user-user clustering for updating individual preferences. This allows VoCoG to account for complex updates from user conversation, while outputting relevant recommendations. \\
\textbf{Conversation Analysis:} There has been extensive work in analyzing natural user conversation. The existing methods describe method for language parsing \cite{ChenManning2014neural, zhu2013fast}, text tagging \cite{ramshaw1995text} and entity recognition \cite{manning2014stanford}. There has also been considerable work in sentiment analysis \cite{bird2006nltk, loria2013text} and intention mining \cite{liu2005opinion, Liu2015ARS} from text. Further, Mikolov \textit{et al} \cite{mikolov2013distributed} have looked into robust semantic representation of words. Commercially, applications like \textit{luis.ai} provide services for entity and intent extraction. VoCoG requires comprehensive parsing of conversation data, including entity extraction, sentiment analysis as well as parsing direct/indirect references in the conversation sequence. Prior work do not address this sufficiently. Hence, we build upon the existing work to analyze user conversation, and extract the required information.\\
\textbf{User-User Interaction Modeling:} There has been work in group behavior modeling in social networks \cite{sherchan2013survey}. However, the area of small group conversation is relatively unexplored. Prior work has looked the problems of conflict resolution \cite{pesarin2012conversation}, identifying speaker \cite{vinyals2008towards} and addressee \cite{jovanovic2004towards} and modeling face-to-face conversations \cite{wyatt2007privacy}. We address the challenge of conflict modeling in multi-user conversations through a novel user-user graph.

\section{System Implementation} \label{sec:ProposedApproach}
In this section, we will describe the modeling for VoCoG, the proposed intelligent assistant interface. The workflow for the approach is described in Figure \ref{fig:VoCoGWorkflow}. The essential components of the system include an online recommendation algorithm, a module to understand the voice conversation between users and inter-user interaction modeling. Each of the these modules are described in details below.

\subsection{Recommendation System} \label{subsec:Recommend}
VoCoG combines a novel incremental collaborative filtering as well as content filtering-based techniques to arrive at a robust ranking of show preferences for individual users. Thereafter, algorithm to update ratings based upon the user conversations is discussed. Note that how the group recommendations are arrived at, will be described later in Section \ref{subsec:GroupAlgorithm}.

\subsubsection{Movie Database} \label{subsubsec:MovieData}
We used MovieLens \cite{Harper2015MovieLens} dataset for training the recommendation system. The dataset has about $20$ million ratings for $27$K movies by around $0.13$ million users. We pruned out users with rating less than $250$ movies and movies having less than $50$ ratings, leaving around $18$K users, $10$K movies and $10$M ratings. This was done to reduce the movie search space during updating VoCoG recommendations. Moreover, relatively unknown movies, not rated by enough users, would not generated conversation among users.  The dataset was further enriched, through crawling the web, with the genre terms, actors and directors for each of the final $10$K movies. This enriched data was used for training the VoCoG recommendation models. 

\subsubsection{Collaborative Filtering}
We chose to deploy a simple probabilistic latent analysis (probLat)-based method for collaborative filtering, but describe a method inspired by \cite{bresler2014latent} to efficiently incorporate complex feedback from user conversations (Section \ref{subsubsec:IncorpUserPrefer}). For a user $u$, his rating $r$, for a movie $m$, was modeled as a function of $z$ and movie $m$. The latent variable $z$ was introduced to decouple probabilistic dependency between users and the movies ratings. Different user interest groups were captured in $z$ and hence, the rating of a movie for a user was calculated as:
\begin{equation}
P(r|u,m) = \sum_{z} p(r|m,z)P(z|u)
\end{equation}
Each user belonged to a cluster $z$ with probability $P(z|u)$ and distribution of ratings across clusters was given by $p(r|m,z)$. Distribution, $p(r|m,z)$, is modeled as:
\begin{equation}
p(r|m,z) \sim \mathcal{N}(\mu_{m,z}, \sigma_{m,z}^2)
\end{equation}


\textbf{Expectation Maximization Algorithm:}
For training the probLat model on the MovieLens dataset, an EM algorithm \cite{hofmann2001unsupervised} was used. The E-step calculated the posterior probability of $z$, given the user $u = u^\prime$, movie $m = m^\prime$ and rating $r = r^\prime$ as:
\begin{equation}
P(z|u,m,r) = \frac{p(r|u,m)P(z|u)}{\sum_{z^\prime}(p(r|u,m)P(z^\prime|u))}
\end{equation}
Once the posterior probabilities were computed, the M-step computed probability of user belonging to different clusters and parameters for the distributions:
\begin{equation}
P(z|u = u^\prime) = \frac{\sum_{u = u^\prime} P(z|u,m,r)} { \sum_{z} \sum_{u = u^\prime} P(z^\prime|u,m,r) }
\end{equation}

\begin{equation}
\mu_{m,z} =  \frac{\sum_{m=m^\prime} rP(z|u,m,r)}{\sum_{m=m^\prime} P(z|u,m,r)}
\end{equation}
 
\begin{equation}
\sigma^2_{m,z} =  \frac{\sum_{m=m^\prime}(r-\mu_{m,z})^2 P(z|u,m,r)}{\sum_{m=m^\prime} P(z|u,m,r)}
\end{equation}
Log-likelihood was used to measure convergence of the algorithm.The algorithm was terminated when change in the log-likelihood went below $2\%$ of the log-likelihood at that step. \\

\subsubsection{Content Filtering}
Content filtering was done through the nearest neighbor approach. Based on the movie rating, $r$ given by user $u$, scores for a genre $g$ and a star $s$ were calculated as follows:
\begin{equation}
genreScore(g) = \frac{\sum_{m \in G_g} r}{\sum_{m \in G_g}1}
\end{equation}
\begin{equation}
starScore[s] = \frac{\sum_{m \in S_s} r}{\sum_{m \in S_s}1}
\end{equation}

where $G_g$ is set of movies containing genre $g$ and $S_s$ is set of movies in which star $s$ has acted. Both $genreScore$ and $starScore$ were normalized with respect to the list of  genres and stars respectively. Content-based score of a movie $m$ for the user $u$ was now calculated by averaging the scores for the genres in the movie and the scores for the stars present in the movie.

\begin{figure}[htp!]
\centering
\includegraphics[width=0.7\columnwidth]{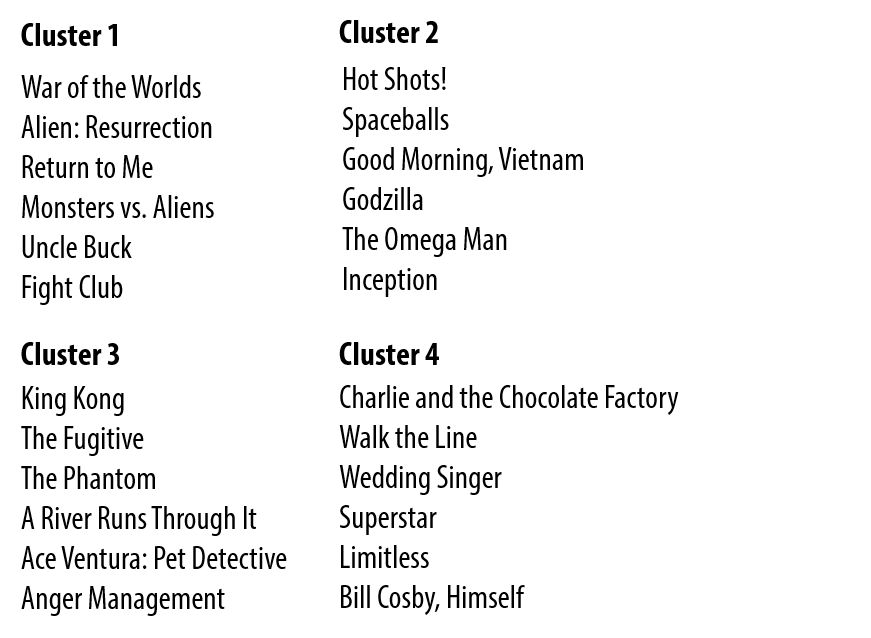}
\caption{Variation in the ranked lists of movies across four clusters.}
\label{fig:MovieRankVary}
\end{figure}

\subsubsection{Incorporating user preferences} \label{subsubsec:IncorpUserPrefer}
\begin{figure}[htp!]
\includegraphics[width=1.0\columnwidth]{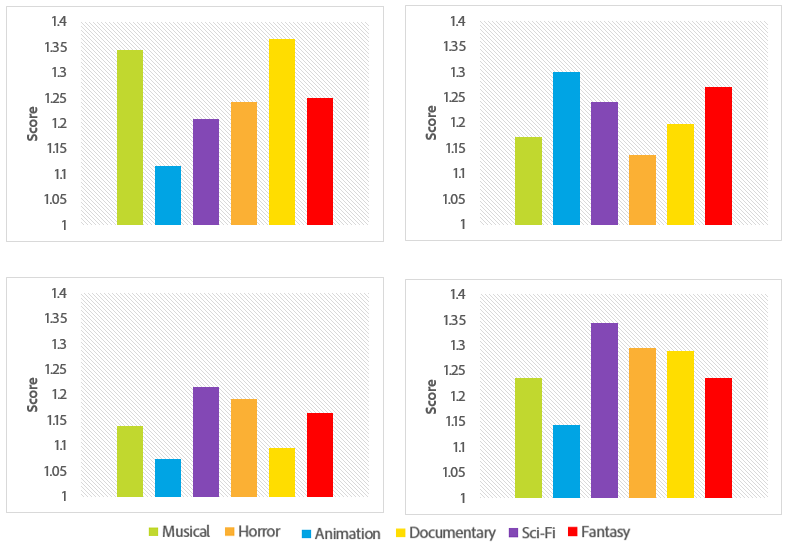}
\caption{Variation of Scores of six genres \textbf{viz.} musical, horror, documentary, sci-fi and fantasy across four clusters.}
\label{fig:GenreScoreVary}
\end{figure}

Here we describe the model for updating recommendations based upon conversation-based feedback. The model can account for feedback for movies, stars or genres from the user. \\
\textbf{1. Ranked Movie List:} Once the probLat model was trained, list of movies was ranked in the descending order of $p(r|m,z)$ for each user interest group, $z$. The mean rating $\mu_{m,z}$ and variance $\sigma_{m,z}^2$ of movie, $m$ varied with cluster $z$. Hence, as shown in the Figure \ref{fig:MovieRankVary}, each cluster $z$ has different movies at the top. The top $500$ movies from each cluster were used for the next step.\\
\textbf{2. Calculating Genre Scores:} For each cluster $z$, scores for different genre terms were calculated by averaging the predicted rating for movies containing the genre term, present in the ranked list. The list of genres was created from the tagged MovieLens data. The genre terms were then ranked in the descending order of scores for each cluster $z$ to get a cluster specific ranked list, $GenreList_{z}$. Figure \ref{fig:GenreScoreVary} shows the variation of genre scores across clusters. The cluster-specific genre scoring was used for updating user preferences.\\
\textbf{3. Incorporating feedback using $P(z|u)$:} We exploited the difference in movie or genre preference across clusters to incorporate conversation feedback, through modifying interest group probability, $P(z|u)$. Different distributions of $P(z|u)$ led to generation of different movies as recommendations from the model. We extracted keywords like genre, movie names, stars from the user conversations, along with attached sentiment as described in Section \ref{subsec:AnalyzeConver}. Here, we describe how to update using the genre preference of the user, but it can be extended to movies or stars as well. 

Let $(g,s)$ be (\textit{genre term, sentiment value}) pair extracted for a particular user conversation. For example, for conversation like "Right now, I am in mood for action movies", the pair would be (\textit{action}, $1$). Then, $P(z|u) = P_{current}(z|u)$ is updated as follows:
\begin{equation}
P_{updated}(z|u) =  P_{current}(z|u)e^{factor}
\end{equation}

where, 
\begin{equation}
factor = \alpha  \times s \frac{N_{z}-R_{g,z}}{N_g},
\end{equation}

$N_g \text{ = total number of genre terms in the list } GenreList_{z},\\ R_{g,z} \text{ = rank of genre } g \text{ in list } GenreList_{z} , \\s \text{ = extracted sentiment and },\\ \alpha \text{ is a hyperparameter.}$

The value of the factor lied between $(-1,1)$, and the exponential ensured that updates can be done serially. The update worked as follows: if a genre, like \textit{action}, ranked higher in clusters $1$, $3$ and $6$ than others and user $u$ expresses a positive sentiment about it, then the probability of user being in cluster $1$, $3$ and $6$ will be increased. Higher the rank of \textit{action} in a cluster, greater will be the update factor for the cluster. Similar equations were used to update $P(z|u)$ using movie and stars keywords. This was repeated for each extracted keyword-sentiment pair. $P(z|u)$ is normalized after all the terms have been processed.\\
\textbf{4. Updating content filtering}: 
We updated content based preference based on the input from conversation for genre preference as follows. Other terms can be similarly taken care of.
\begin{equation}
genreScore[g] = genreScore[g] e^{\alpha \times s}
\end{equation}

\subsubsection{Implementation and Results}
For implementation, the number of clusters in probLat model was taken to be $16$ and the hyperparameter $\alpha$ was set to $0.5$ empirically. The final recommendations were arrived by a simple average of probLat and content filtering scores. The probModel was also compared with different methods in literature (Table \ref{tab:pLSAResults}). For the testing, rating of a random movie among the movies rated by each user was removed. The model was then trained on the reduced data set and tested on the movies removed from the dataset. It can be seen that the probLat method shows comparable performance with some of the previous methods.  

\begin{table}[htp!]
	\caption{Comparison of recommendation accuracy for the probLat model.}
	\label{tab:pLSAResults}
	\begin{center}
	\begin{tabular}{@{}lll@{}}
		\toprule
		Method & Mean Absolute Error  \\ \midrule
		probLat (clusters = $16$) & $0.658$   \\
		SVD \cite{mehta2007robust}  & $0.67$  \\ 
		k-NN \cite{mehta2007robust} & $0.79$ \\ \hline
	\end{tabular}
		\end{center}

\end{table}

\subsection{Natural Language Understanding} \label{subsec:AnalyzeConver}
It is challenging to process human language, more so when the people are conversing. For incorporating non-intrusive feedback in VoCoG, it was required to design a workflow which could update the viewer preferences solely based on their conversation. For this, we broke down the entire complex conversation to simpler keyword - sentiment pairs, which could then be used to update user preferences as discussed in Section \ref{subsec:Recommend}. The keywords included movie mentions, named entities like stars or directors and mention of genre terms. The process is discussed in described in details below. 

\subsubsection{Speech-to-text Conversion} The user conversations were first converted to text using existing APIs \cite{schalkwyk2010your}. Though the accuracy of speech-to-text APIs have increased considerably, there are accompanying challenges in further processing as described below. The conversations were analyzed sentence-wise.

\subsubsection{Conversation Database} \label{subsubsec:ConverseData}
A database of user conversations (MovieForum) was curated from a movie discussion forum (movieforums.com). The dataset created had $10$ different threads and an average of $180$ comments in each thread and each thread involved $6$ users on an average. The conversations were manually labeled with movie, genre and actor mentions. We also further tagged the conversation with the mentions of user, who are involved in the discussion. These tagging can be either direct mentions of a user/star or indirectly through the use of pronouns, etc. Each of the conversations were further labeled manually with a sentiment value ($-1$, $0$ or $1$).

For the purpose of evaluating different tagging and sentiment detection approaches, we used $50:50$ train/test split of the corresponding dataset. The hyper-parameters were trained using $5$-fold cross validation scheme. In the cases where no training was required, full dataset was used for evaluation.

\subsubsection{Sentence-Level Keyword Extraction}
We describe below the proposed methods for extraction of different types of keywords, like genre, movie, actor.\\
\textbf{1. Genre Terms:} Most of the genre terms in the curated MovieLens database from Section \ref{subsec:Recommend} (like action, drama) were single words. So, the genre terms were extracted using simple word search. The look-up list of genres was compiled from the movie database. The method gave an F-score of around $0.9$ on the MovieForum database.\\
\textbf{2. Movies Terms:} Movie names were more complex like "One Flew over Cuckoo's nest". Also, there was a comprehensive movie list to search for (around 10k for our MovieLens dataset). Hence, a two-step process was used for extraction: \\
\textbf{a. Movie Tagging:} Alternate methods of tagging potential movie mentions were compared for this purpose. \\
\textit{Baseline approach:} The existing, state-of-the-art POS tagging method \cite{ChenManning2014neural} was used to detect nouns from the sentences, and then use the detected parts as the tagging for movie mentions. \\ 
\textit{Learning-based approach:} The training data from the MovieForum data was IOB-tagged \cite{ramshaw1995text}. The features used for training gradient-boosted classifier included the POS tags of current as well as that for words in a window of length $5$ around the word, position of the word (e.g., first or last word), a vector representation of word provided by word2vec model \cite{mikolov2013distributed} and if the word is among top $3000$ most frequent word in movie name list (MovieLens data). The output of this classifier was smoothened using an HMM-based sequence analyzer, trained on the MovieForum data with I,O,B as hidden states. This was done to weed out some of the unlikely classifications done by the classifier. The method overcame challenges of unreliable capitalization and could detect long names as well. The performance of the tagging approaches are summarized in Table \ref{tab:MovieTaggingResult}.
\begin{table}[htp!]
	\caption{Results for movie tagging detection.}
	\label{tab:MovieTaggingResult}
	\begin{center}
	\begin{tabular}{@{}lll@{}}
		\toprule
		& Precision & Recall \\ \midrule
		Baseline \cite{ChenManning2014neural}    & $0.61$     & $0.69$  \\
		Proposed Approach   & $\mathbf{0.85}$     & $\mathbf{0.75}$  \\ \hline
	\end{tabular}
	\end{center}
\end{table}

\textbf{b. Movie Name Search:} The tagged output was then matched with the movie names in the MovieLens database using a string search, based on the Levenshtien distance measure. The top $10$ ranked movies were then re-ranked on the basis of the context of the conversation. Context included genre or actor detected in the previous $10$ conversation. The scores for the movies related to these mentions were increased, and then ranked accordingly. Table \ref{tab:MovieNameResult} shows the overall performance of the movie extractor.
\begin{table}[htp!]
	\caption{Results for movie name recognition.}
	\label{tab:MovieNameResult}
	\begin{center}
	\begin{tabular}{@{}lll@{}}
		\toprule
		& Precision & Recall \\ \midrule
		Baseline \cite{ChenManning2014neural}      & $0.52$     & $0.72$  \\
		Proposed Approach   & $\mathbf{0.76}$     & $\mathbf{0.78}$ \\ \hline
	\end{tabular}
	\end{center}
\end{table}

\textbf{3. Movie Stars Terms:} Movie stars were detected following the method used for movies. The stars tagging method was compared to the Stanford name-entity tagger \cite{manning2014stanford}. It can be seen in Table \ref{tab:StarNameResult} that the proposed approach outperforms the recall of the Stanford tagger, with only a small decrease in precision. 
\begin{table}[htp!]
	\caption{Results for star name recognition.}
	\label{tab:StarNameResult}

	\begin{center}
	\begin{tabular}{@{}lll@{}}
		\toprule
		& Precision & Recall \\ \midrule
		Stanford Name Tagger \cite{manning2014stanford} & $\mathbf{0.98}$      & $0.49$   \\
		Proposed Approach         & $0.91$      & $\mathbf{0.89}$  \\
		\hline
	\end{tabular}
	\end{center}
\end{table}

\textbf{4. Indirect references:} References to a movie or a star using determiners like it or him/her were attached to the last mention of a movie or a star, detected from the conversation. 

\subsubsection{Sentence-Level Sentiment Analysis}
The existing sentiment analysis methods were found insufficient for our case. They did not classify sentiment for intent well, e.g. "We should be watching Inception" was classified as a neutral sentiment.  They also did not take care of sentences framed as questions, e.g. "Why shouldn't we watch inception?". The baseline approaches assigned negative sentiment to the sentence. There were cases like the negative sentiment being assigned due to the movie name itself, e.g. "Let us watch Wrong Turn". 
Hence, a modified sentiment analyzer was trained. Features included - 1. if the sentence is a question or not, 2. presence of words indicating intention , positive or negative words \cite{liu2005opinion}, 3. average representation of sentiment and intention words given by word2vec model \cite{mikolov2013distributed} and 4. scores of existing sentiment classifiers \cite{bird2006nltk, loria2013text}. Also to avoid the problem of a keyword (movie or actor) altering the sentiment, the positive or negative keywords were removed. The performance comparison of the developed sentiment analyzer for the MovieForum data is provided in Table \ref{tab:SentimentResults}.

\begin{table}[htp!]
	\caption{Results for sentiment analysis on the MovieForum dataset.}
	\label{tab:SentimentResults}

	\begin{center}
	\begin{tabular}{@{}lc@{}}
		\toprule
		& Accuracy (\%) \\ \midrule
		NLTK Sentiment Classifier \cite{bird2006nltk} & $56 \%$   \\
		Text Blob Classifier \cite{loria2013text}     & $43 \%$   \\
		Proposed Approach              & $\mathbf{79 \%}$       \\ \hline   
	\end{tabular}
	\end{center}
\end{table}

\subsubsection{Keyword-Sentiment Pairing}
The last step was to attach sentiment to the extracted keywords. The direct approach was pair the sentence sentiment with the corresponding keywords. However, in conversations, people can mention multiple movies in a sentence, with contrasting sentiment. Hence, we used a set of linguistic-based rules to improve the pairing, as described below. 
\begin{itemize}
\item The sentence was parsed using a constituency parser \cite{zhu2013fast} and a set of rules were created to attach the sentiment to the keyword.
\item  A total of 20 rules were created for comparative words like "but", "and", "or", "yet", "although", "both ... and", "instead", "as ... as", "than". \textit{E.g.} the rule for "but" was: In the constituency parse tree, if the parent of "but" conjunction is a noun phrase, attach the reverse sentiment of the part containing the verb phrase to the part which does not contain the verb phrase.
\end{itemize}
The final results for keyword-sentiment pairing are shown in Table\ref{tab:KeywordSentimentPair}. 
\begin{table}[htp!]
	\caption{Comparison of results for keyword-sentiment pairing.}
	\label{tab:KeywordSentimentPair}

	\begin{center}
	\begin{tabular}{@{}lll@{}}
		\toprule
		& Precision & Recall \\ \midrule
		Direct Pairing   & $0.52$      & $0.59$   \\
		Modified Pairing & $\textbf{0.68}$      & $\textbf{0.67}$  \\ \hline
	\end{tabular}
	\end{center}
\end{table}

\subsection{Inter-User Influence Modeling} \label{subsec:GroupAlgorithm}
In this section we describe the modeling of inter-user influence from conversation. We explain how the ratings of users vary due to agreement or conflict during the conversation. We create a user-user graph, based upon related work in social networks \cite{sherchan2013survey}. The algorithm assumes the knowledge of the user names of people present in the conversation, and takes the keyword-sentiment pairs, extracted in Section \ref{subsec:AnalyzeConver}, as the input.\\
\textbf{1. Dependency Parsing: }First parts of speech tags (POS tags) were detected using dependency parsing \cite{ChenManning2014neural}, and which were then used to detect the subject of conversation. The keywords like movie, actors and the corresponding sentiment were extracted as explained in Section \ref{subsec:AnalyzeConver}. The detected (subject, keyword, sentiment) tuples were outputted.\\
\textbf{2. Keyword pruning: }In conversation, there would be cases in which references to a movie or star can not be linked to another user. An example would be "I want to watch The Prestige". "I-The Prestige" would be the user-keyword pair obtained from this, but in case The Prestige was not referred to before by any user, it would not convey agreement or disagreement with any other user. These keywords were pruned out. \\
\textbf{3. Inter-User Sentiment: }We now find the expressed sentiment for interaction between users. There are two possible cases here: 
\begin{itemize}
\item If the subject was not detected, the user who last used the particular keyword was taken as the referred user. The agreement or disagreement (i.e. the sentiment of interaction) is determined by whether their expressed sentiments matched or not.
\item If the keywords and subject were not detected from the sentence, then the following method was used. We assumed that people talking about what someone else has talked about, tend to bring up similar topics. Hence, we find the overlap of noun words between the sentences of the user as well as recent sentences spoken before. The user who last spoke the maximum overlapping sentence was assigned as the user referred. In case there was no overlap, the speaker of the sentence previous to the current one was taken to be the referred user.
\end{itemize}
\textbf{5. User-User Influence Graph: }The sentiment for ordered user pair $(i,j)$ from conversation was used to update the graph. The weight $w_{i,j}$ was assigned to be the extracted sentiment. Note that the graph is not symmetrical, as user $j$ agreeing or disagreeing with user $i$ changes $w_{i,j}$, but not $w_{j,i}$. For multiple conversations, the sentiment for each one was added to the corresponding weight value.\\
\textbf{6. User Rating Matrix update: }The rating, $r_{m,i}$ for a movie, $m$ by user, $i$ was updated using the user-user influence graph as follows:
\begin{equation}
r_{m,i}=r_{m,i}+\sum_{j \neq i,w_{ji}>0}\frac{w_{ji}}{\sum_{k \neq i,w_{ki}>0}w_{ki} +\alpha} \times (\frac{r_{m,j}+r_{m,i}}{2}-r_{m,i}),
\end{equation}
where $\alpha$ is a regularization parameter. In our tests, $\alpha$ was set to be the number of users. The update brought the rating of users in agreement closer together, so as to arrive at consensus quicker. Negative weight edges in the graph were not used in the update. However, the negative weights were maintained so that users, who were in prior disagreement, must come to agreement before the correspond edge weight to be taken into account.\\
\textbf{7. Limitations: }Our subject analysis method may fail in case of complex movie names, for example "Who Is Harry Kellerman and Why Is He Saying Those Terrible Things About Me?". If this movie is part of a sentence, naturally "He" will be detected as a subject, as well as a pronoun, and this will lead to a result that suggests the presence of an inter-user interaction, although there may not be. 
\begin{figure}
\centering
\includegraphics[scale = 0.45]{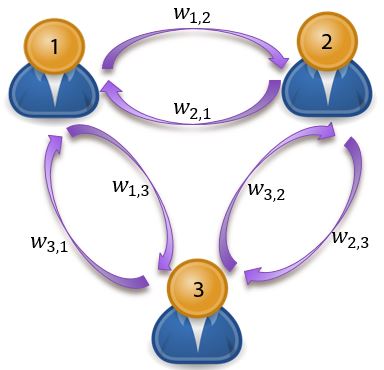}
\caption{User-user graph for modeling inter-user interactions.}
\label{fig:UserUserGraph}
\end{figure}
\subsubsection{Results}
The user influence modeling system showed good performance on the MovieForum dataset. For a set of about $40$ users in the database with agreements, disagreements or neutral exchanges between any two users, the algorithm had $precision = 0.72$ and $recall = 0.61$.
\begin{table*}[t]
\caption{Results of the conducted survey of VoCoG, the proposed collaborative viewing system on the Likert Scale of $1-5$, with each cell denoting the fraction of responses. On all the components except the response time, more than $50 \%$ participants showed agreement or strong agreement. \textit{p-value} for Wilcoxon test with hypothesis of being greater than rating of $3$ is shown in the last table.}
\label{tab:SurveyResults}
	\begin{center}
	\begin{tabular}{lcccccc} \toprule
 \textbf{Questions} & \textbf{Strongly Disagree(1)} & \textbf{Disagree(2)} & \textbf{Neutral(3)} & \textbf{Agree(4)} & \textbf{Strongly Agree(5)} & \textbf{p-value}  \\ \midrule
		Overall VoCoG provided good experience    & $0$ & $0$ & $0.2$ & $0.6$ & $0.2$ & $1.62e-5^*$ \\
		Final recommendations were good & $0$ & $0.1$ & $0.3$ & $0.5$ & $0.1$ & $4.93e-3^*$      \\
		Updates to recommendations were appropriate & $0.0$ & $0.1$ & $0.3$ & $0.3$  & $0.3$ & $2.13e-4^*$  \\
		System took care of your preferences & $0.0$ & $0.1$ & $0.2$ & $0.4$ & $0.3$  & $1.74e-4^*$   \\
		Response time of the system was fast enough & $0.1$ & $0.2$ & $0.4$ & $0.3$ & $0$ & $0.68$ \\
		System was non-intrusive & $0$ & $0$  & $0.2$ & $0.5$ & $0.3$ & $5.09e-6^*$   \\ \hline
	\end{tabular}
	\end{center}
\end{table*}

\begin{figure*}
\centering
\includegraphics[scale=0.4]{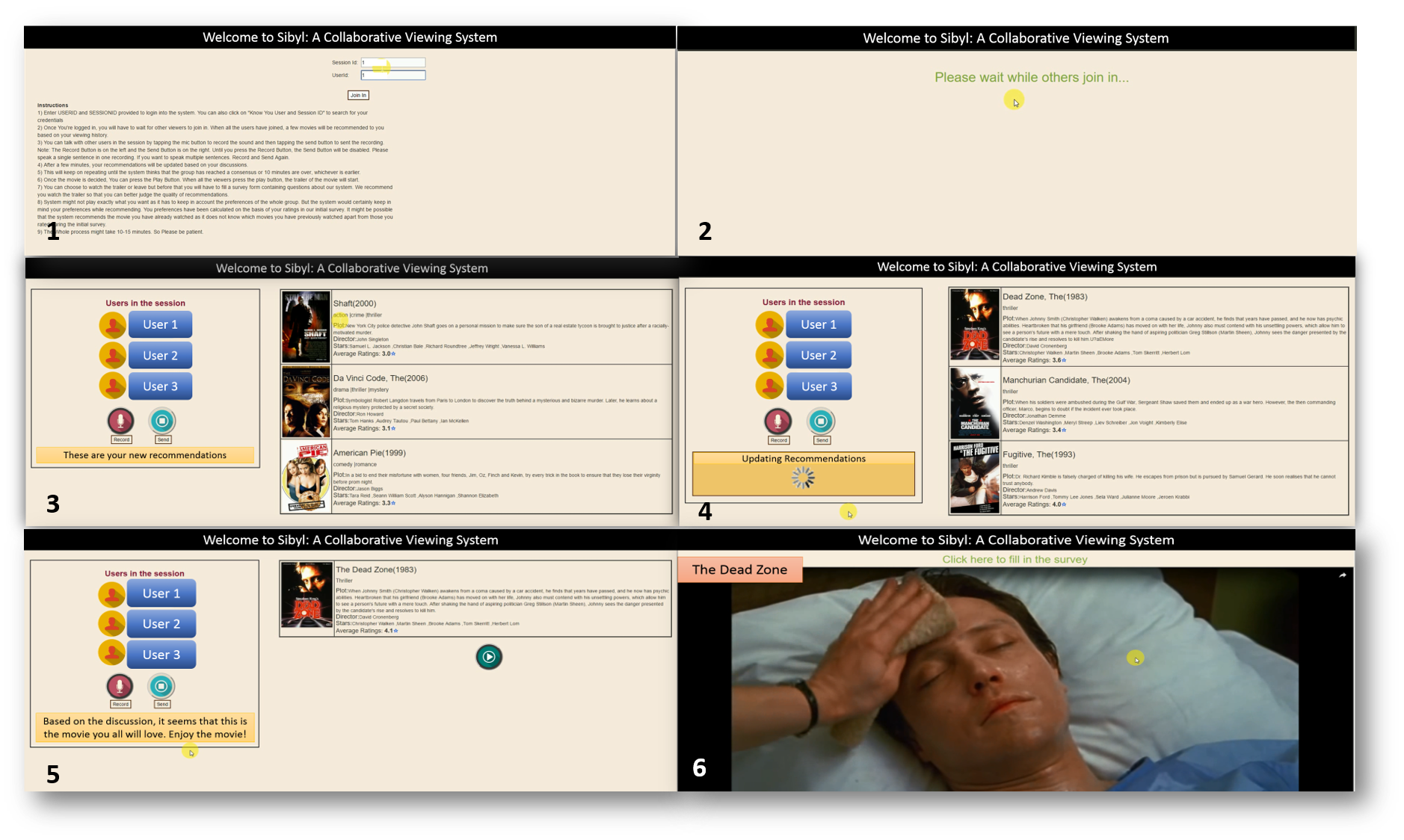}
\caption{Snapshots of the interface of a prototype of the proposed system.}
\label{fig:VoCoGWorkingDemo}
\end{figure*}

\subsection{Group Consensus Function}
Finally, to arrive at the recommendations for the whole group, a group consensus function was used. A variety of group consensus functions like maximum pleasure, average satisfaction , least misery, etc. have been explored in the group recommendation literature \cite{masthoff2011group,mccarthy2006group}.\\
\textbf{1. Average without misery function:} We used "average without misery" function for our case. This function first eliminates movies on the basis of "misery", i.e. if any user has rated a movie below a threshold, then the movie is eliminated. For the surviving movies, average rating is computed for each movie, based on which the top movies are decided. In our experiments, the threshold for misery was decided empirically. The final decision was taken using a weighted average rating computation. The weights were decided upon by user behavior in the conversation, like sentences spoken and users influenced.\\
\textbf{2. Consensus Detection:} The system decided whether the users have reached a consensus by comparing the top rated movie with the lower ranked ones. If the overall rating of the movie for the group exceeded the next movie in the list by a specified threshold (set to $1.5$ times in our experiment), then the consensus was deemed to have been reached.

\section{Working Prototype} \label{sec:SysUserInterface}
Figure \ref{fig:VoCoGWorkingDemo} shows snapshots of a working prototype of the system. As shown in the figure, first the users are asked to login into the system. VoCoG waits till all the users have joined the session, to enable a synchronized experience. Once all the users have logged in, VoCoG generates initial recommendations based on the users' previous histories, and outputs a voice message as well as a text on the screen. In the prototype, the number of movie options shown was kept at $3$ so as to generate conversation about each option. 

 The users can then converse among themselves using "Record" and "Send" buttons. This is similar to the interfaces in many voice-based assistant systems. The interface also helps in sequencing the user conversation seamlessly. The sentences spoken by the users are sent to other users and also the back-end server in real time. The users are represented by avatars, and as they speak, the corresponding avatars light up.

After a fixed time interval, VoCoG refreshes the recommendations and an "Updating recommendations" message is played as well as shown on the screen. The next set of recommendations are then displayed, which is based on processing the user conversation following the method described in Section \ref{sec:ProposedApproach}. The process continues until the consensus is detected, as shown in the figure, where users have converged upon the movie "The Dead Zone". The movie is then played when all the users click the video icon. In case the consensus is not reached after five rounds of updates, the top rated movie is shown as the final output.

\section{System Evaluation} \label{sec:SysEvaluation}
For a comprehensive evaluation of the proposed system, VoCoG was made to interact with users. About $45$ people ($10$ females) were involved in a survey to judge the performance of VoCoG. The participants in the survey were drawn from the age group of $20-30$, and had varied movie watching preferences. The system was then measured on different parameters following the methods for user-centric recommender system evaluation \cite{knijnenburg2012explain, pu2011user}. 
 
\subsection{Survey Design}
The survey was conducted as follows: \\
\textbf{Phase 1:} In the first phase the participants were asked to rate a set of popular movies from the MovieLens database. Movies to be rated were chosen to be representative of different genres. These ratings were used to train the model combined with the movie dataset (Section \ref{subsubsec:MovieData}). Each subject rated $12$ movies on an average. This gave data to VoCoG to create an initial profile of the subject. The subjects were also asked if they were frequent movie watcher (more than 2 times a week), and if they are usually active in conversation.\\
\textbf{Phase 2:} In the second phase these subjects were grouped in groups of $3$, and $15$ groups were formed.  These groups were then called upon to interact with the system. Thereafter, they rated the system on six parameters, as shown in Table \ref{tab:SurveyResults}, on a Likert scale of $1-5$ , where $1$ represents the worst rating and $5$ the best. Arrangements were made to have an environment identical to the one which the viewers would experience in a remote collaborative viewing implementation. All the three viewers were made to sit in different rooms and could interacting only through the system. VoCoG listened to their conversations, and updated the recommendations periodically. \\
\textbf{Questionnaire:} After the interaction, the participants were made to fill a questionnaire. Here, they rated the different aspects of interaction with the system (shown in Table \ref{tab:SurveyResults}) on a Likert scale of $1$ (Strongly disagree) - $5$ (Strongly Agree). 

\subsection{Survey Analysis}
Here, we analyze different aspects of interaction of participants with VoCoG. \\ 
\textbf{1. Questionnaire Response:} The summary of the questionnaire responses is shown in Table \ref{tab:SurveyResults}. As can be seen, VoCoG received strongly positive response (more than $60 \%$ participants agreed or strongly agreed) for all the parameters (recommendation quality, interactivity, non-intrusiveness) except response time. This is because VoCoG searches through a large movie dataset for recommendations. We hope to improve the system response time in future implementations. \\
\textbf{2. Conversation analysis:} Table \ref{tab:EntityStats} shows the statistics of average mentions of different entities in the survey. It can be seen that the participants conversed the most about movies, followed by genres and actors/stars. There were also considerable agreements/disagreements between the participants while interacting. Overall they participated well in the survey, with number of sentences spoken per update being around $8$.\\
\textbf{3. Group Recommendation response:} Table \ref{tab:EntityStats} also shows the statistics for user responses to the recommendations provided by VoCoG. As the users can provide feedback through conversation, different aspects of the response are required to be captured (different from click-based systems). As shown in the Table \ref{tab:EntityStats}, about $1.4$ movie mentions out of the total $3.1$ average mentions per update cycle were regarding the recommended movies. Overall on an average $2.1$ out of $3$ movies were unique per update. This shows that while the users discussed the recommended movies, they also looked out for diverse recommendations. The statistics for genre term mentions ($1.5$ out of $2.5$ on an average were from recommended list) indicate that the users expressed more conveniently in terms of their genre choices. Actors and directors were mentioned only few times. Also, VoCoG was able to reach consensus for only $3$ out of $15$ groups. This calls for a need for better modeling for group consensus and understanding user dynamics. We intend to study these as future directions to the work.\\
\textbf{4. Variations due to user differences:} We also studied how the nature of participants, \textit{viz.} frequent/non-frequent and active/non-active (as collected in the Phase 1) affected their interaction with the system. As shown in Table \ref{tab:ExpertStats}, frequent and active participants rated VoCoG highly on overall experience, but there were some lower ratings by non-frequent and non-active participants.
\begin{table}[htp!]
\caption{Analysis of average mentions of different entities in the interaction of participants with VoCoG, between consecutive recommendation updates.}
\label{tab:EntityStats}	
	\begin{center}
	\begin{tabular}{lc} \toprule
	\textbf{Entities} & \textbf{Avg. number per update} \\ \midrule
	Sentences spoken & $7.8$ \\
	Movie mentions & $3.1$ \\
	Actors/Directors mentioned & $1.8$ \\
	Genre Terms &  $2.5$ \\
	Unique movies recommended & $2.1$ \\
	Recommended Movies mentions & $1.4$ \\
	Recommended Genre mentions & $1.5$ \\
	Recommended Actors mentions & $0.4$ \\
	User Agreement/Disagreement & $2.1$ \\ \hline
	\end{tabular}
	\end{center}
\end{table}
\begin{table}[htp!]
\caption{Ratings statistics for different participant groups (frequent/non-frequent movie watchers, active/non-active in conversation) on the overall experience with VoCoG being good.}
\label{tab:ExpertStats}
	\begin{center}
	\begin{tabular}{lccccc} \toprule
	\textbf{Participant} & \textbf{SD} & \textbf{D} & \textbf{N} & \textbf{A} & \textbf{SA} \\ \midrule
	Frequent, Active & $0.0$ & $0.0$ & $0.0$ & $0.7$ & $0.3$ \\
	Freuent, Non-Active & $0.0$ & $0.0$ & $0.2$ & $0.6$ & $0.2$ \\
	Non-frequent, Active & $0.0$ & $0.1$ & $0.1$ & $0.6$ & $0.2$ \\
	Non-frequent, Non-active & $0.0$ & $0.2$ & $0.2$ & $0.6$ & $0.0$ \\
	\hline
	\end{tabular}
	\end{center}
	
\end{table}

\section{Conclusion and Future Directions} \label{sec:Conclusions}
In this paper, we have described framework for VoCoG, an intelligent, non-intrusive interface for collaborative group-viewing experience. We have described the technology behind each components of VoCoG, \textit{viz.} an online recommendation system, a robust conversation analyzer and a user-user interaction modeling algorithm. 

In the future, we plan to optimize the system for an efficient response time. We also need to expand the scope of the algorithms to update user preferences beyond the session, for longer-term viewing experience optimization and incorporate better features for user dynamics and consensus modeling. We further plan to incorporate a richer GUI, using avatars and augmented sound to improve the experience.

\balance{}

\bibliographystyle{IEEEtran}
\bibliography{paper_biblio}

\begin{thebibliography}{10}
\providecommand{\url}[1]{#1}
\csname url@samestyle\endcsname
\providecommand{\newblock}{\relax}
\providecommand{\bibinfo}[2]{#2}
\providecommand{\BIBentrySTDinterwordspacing}{\spaceskip=0pt\relax}
\providecommand{\BIBentryALTinterwordstretchfactor}{4}
\providecommand{\BIBentryALTinterwordspacing}{\spaceskip=\fontdimen2\font plus
\BIBentryALTinterwordstretchfactor\fontdimen3\font minus
  \fontdimen4\font\relax}
\providecommand{\BIBforeignlanguage}[2]{{%
\expandafter\ifx\csname l@#1\endcsname\relax
\typeout{** WARNING: IEEEtran.bst: No hyphenation pattern has been}%
\typeout{** loaded for the language `#1'. Using the pattern for}%
\typeout{** the default language instead.}%
\else
\language=\csname l@#1\endcsname
\fi
#2}}
\providecommand{\BIBdecl}{\relax}
\BIBdecl

\bibitem{blascovich2011infinite}
J.~Blascovich and J.~Bailenson, \emph{Infinite reality: Avatars, eternal life,
  new worlds, and the dawn of the virtual revolution}.\hskip 1em plus 0.5em
  minus 0.4em\relax William Morrow \& Co, 2011.

\bibitem{Cesar2009FTIH}
P.~Cesar and K.~Chorianopoulos, ``The evolution of tv systems, content, and
  users toward interactivity,'' \emph{Foundations and Trends in Human-Computer
  Interaction}, vol.~2, no.~4, pp. 373--95, Apr 2009.

\bibitem{nathan2008collaboratv}
M.~Nathan, C.~Harrison, S.~Yarosh, L.~Terveen, L.~Stead, and B.~Amento,
  ``Collaboratv: making television viewing social again,'' in \emph{Proceedings
  of the 1st international conference on Designing interactive user experiences
  for TV and video}.\hskip 1em plus 0.5em minus 0.4em\relax ACM, 2008, pp.
  85--94.

\bibitem{lull1980social}
J.~Lull, ``The social uses of television,'' \emph{Human communication
  research}, vol.~6, no.~3, pp. 197--209, 1980.

\bibitem{webster1982impact}
J.~G. Webster and J.~J. Wakshlag, ``The impact of group viewing on patterns of
  television program choice,'' \emph{Journal of Broadcasting \& Electronic
  Media}, vol.~26, no.~1, pp. 445--455, 1982.

\bibitem{ducheneaut2008social}
N.~Ducheneaut, R.~J. Moore, L.~Oehlberg, J.~D. Thornton, and E.~Nickell,
  ``Social tv: Designing for distributed, sociable television viewing,''
  \emph{Intl. Journal of Human-Computer Interaction}, vol.~24, no.~2, pp.
  136--154, 2008.

\bibitem{Ko2016TOCHI}
M.~Ko, S.~Choi, J.~Lee, U.~Lee, and A.~Segev, ``Understanding mass interactions
  in online sports viewing: Chatting motives and usage patterns,'' \emph{ACM
  Trans. Comput.-Hum. Interact.}, vol.~23, no.~1, pp. 6:1--6:27, Jan. 2016.

\bibitem{McGill2016TOCHI}
M.~McGill, J.~H. Williamson, and S.~Brewster, ``Examining the role of smart tvs
  and vr hmds in synchronous at-a-distance media consumption,'' \emph{ACM
  Trans. Comput.-Hum. Interact.}, vol.~23, no.~5, pp. 33:1--33:57, Nov. 2016.

\bibitem{steuer1992define}
J.~Steuer, ``Defining virtual reality: Dimensions determining telepresence,''
  \emph{Journal of communication}, vol.~42, no.~4, pp. 73--93, 1992.

\bibitem{sanchez2005presence}
M.~V. Sanchez-Vives and M.~Slater, ``From presence to consciousness through
  virtual reality,'' \emph{Nature Reviews Neuroscience}, vol.~6, no.~4, pp.
  332--339, 2005.

\bibitem{Converse2016KDD}
K.~Christakopoulou, F.~Radlinski, and K.~Hofmann, ``Towards conversational
  recommender systems,'' in \emph{Proceedings of the 22Nd ACM SIGKDD
  International Conference on Knowledge Discovery and Data Mining}, ser. KDD
  '16.\hskip 1em plus 0.5em minus 0.4em\relax New York, NY, USA: ACM, 2016, pp.
  815--824.

\bibitem{chen2012critique}
L.~Chen and P.~Pu, ``Critiquing-based recommenders: survey and emerging
  trends,'' \emph{User Modeling and User-Adapted Interaction}, vol.~22, no.~1,
  pp. 125--150, 2012.

\bibitem{bresler2014latent}
G.~Bresler, G.~H. Chen, and D.~Shah, ``A latent source model for online
  collaborative filtering,'' in \emph{Advances in Neural Information Processing
  Systems}, 2014, pp. 3347--3355.

\bibitem{zhao2013interactive}
X.~Zhao, W.~Zhang, and J.~Wang, ``Interactive collaborative filtering,'' in
  \emph{Proceedings of the 22nd ACM international conference on Conference on
  information \& knowledge management}.\hskip 1em plus 0.5em minus 0.4em\relax
  ACM, 2013, pp. 1411--1420.

\bibitem{kawale2015efficient}
J.~Kawale, H.~H. Bui, B.~Kveton, L.~Tran-Thanh, and S.~Chawla, ``Efficient
  thompson sampling for online￼ matrix-factorization recommendation,'' in
  \emph{Advances in Neural Information Processing Systems}, 2015, pp.
  1297--1305.

\bibitem{sherchan2013survey}
W.~Sherchan, S.~Nepal, and C.~Paris, ``A survey of trust in social networks,''
  \emph{ACM Computing Surveys (CSUR)}, vol.~45, no.~4, p.~47, 2013.

\bibitem{masthoff2011group}
J.~Masthoff, ``Group recommender systems: Combining individual models,'' in
  \emph{Recommender systems handbook}.\hskip 1em plus 0.5em minus 0.4em\relax
  Springer, 2011, pp. 677--702.

\bibitem{mccarthy2006group}
K.~McCarthy, M.~Salam{\'o}, L.~Coyle, L.~McGinty, B.~Smyth, and P.~Nixon,
  ``Group recommender systems: a critiquing based approach,'' in
  \emph{Proceedings of the 11th international conference on Intelligent user
  interfaces}.\hskip 1em plus 0.5em minus 0.4em\relax ACM, 2006, pp. 267--269.

\bibitem{carlsson1993dive}
C.~Carlsson and O.~Hagsand, ``Dive a multi-user virtual reality system,'' in
  \emph{Virtual Reality Annual International Symposium, 1993., 1993
  IEEE}.\hskip 1em plus 0.5em minus 0.4em\relax IEEE, 1993, pp. 394--400.

\bibitem{curtis1994muds}
P.~Curtis and D.~A. Nichols, ``Muds grow up: Social virtual reality in the real
  world,'' in \emph{Compcon Spring'94, Digest of Papers.}\hskip 1em plus 0.5em
  minus 0.4em\relax IEEE, 1994, pp. 193--200.

\bibitem{igarashi2001voice}
T.~Igarashi and J.~F. Hughes, ``Voice as sound: using non-verbal voice input
  for interactive control,'' in \emph{Proceedings of the 14th annual ACM
  symposium on User interface software and technology}.\hskip 1em plus 0.5em
  minus 0.4em\relax ACM, 2001, pp. 155--156.

\bibitem{cohen1995role}
P.~R. Cohen and S.~L. Oviatt, ``The role of voice input for human-machine
  communication,'' \emph{Proceedings of the National Academy of Sciences},
  vol.~92, no.~22, pp. 9921--9927, 1995.

\bibitem{salzman1999model}
M.~C. Salzman, C.~Dede, R.~B. Loftin, and J.~Chen, ``A model for understanding
  how virtual reality aids complex conceptual learning,'' \emph{Presence:
  Teleoperators and Virtual Environments}, vol.~8, no.~3, pp. 293--316, 1999.

\bibitem{becker2016interactive}
V.~Becker, ``Interactive television experience in convergent environment:
  Models, reception and business,'' in \emph{Proceedings of the ACM
  International Conference on Interactive Experiences for TV and Online
  Video}.\hskip 1em plus 0.5em minus 0.4em\relax ACM, 2016, pp. 119--122.

\bibitem{adomavicius2005toward}
G.~Adomavicius and A.~Tuzhilin, ``Toward the next generation of recommender
  systems: A survey of the state-of-the-art and possible extensions,''
  \emph{IEEE transactions on knowledge and data engineering}, vol.~17, no.~6,
  pp. 734--749, 2005.

\bibitem{roy2014exploiting}
S.~B. Roy, S.~Thirumuruganathan, S.~Amer-Yahia, G.~Das, and C.~Yu, ``Exploiting
  group recommendation functions for flexible preferences,'' in \emph{2014 IEEE
  30th International Conference on Data Engineering}.\hskip 1em plus 0.5em
  minus 0.4em\relax IEEE, 2014, pp. 412--423.

\bibitem{Wu2008Recsys}
H.~Wu, Y.~Wang, and X.~Cheng, ``Incremental probabilistic latent semantic
  analysis for automatic question recommendation,'' in \emph{Proceedings of the
  2008 ACM Conference on Recommender Systems}, ser. RecSys '08.\hskip 1em plus
  0.5em minus 0.4em\relax New York, NY, USA: ACM, 2008, pp. 99--106.

\bibitem{stern2009matchbox}
D.~H. Stern, R.~Herbrich, and T.~Graepel, ``Matchbox: large scale online
  bayesian recommendations,'' in \emph{Proceedings of the 18th international
  conference on World wide web}.\hskip 1em plus 0.5em minus 0.4em\relax ACM,
  2009, pp. 111--120.

\bibitem{ChenManning2014neural}
\BIBentryALTinterwordspacing
D.~Chen and C.~Manning, ``{A Fast and Accurate Dependency Parser using Neural
  Networks},'' in \emph{Proceedings of the 2014 Conference on Empirical Methods
  in Natural Language Processing (EMNLP)}.\hskip 1em plus 0.5em minus
  0.4em\relax Association for Computational Linguistics, Oct. 2014, pp.
  740--750. [Online]. Available: \url{http://www.aclweb.org/anthology/D14-1082}
\BIBentrySTDinterwordspacing

\bibitem{zhu2013fast}
M.~Zhu, Y.~Zhang, W.~Chen, M.~Zhang, and J.~Zhu, ``Fast and accurate
  shift-reduce constituent parsing.'' in \emph{ACL (1)}, 2013, pp. 434--443.

\bibitem{ramshaw1995text}
L.~A. Ramshaw and M.~P. Marcus, ``Text chunking using transformation-based
  learning,'' \emph{arXiv preprint cmp-lg/9505040}, 1995.

\bibitem{manning2014stanford}
C.~D. Manning, M.~Surdeanu, J.~Bauer, J.~R. Finkel, S.~Bethard, and
  D.~McClosky, ``The stanford corenlp natural language processing toolkit.'' in
  \emph{ACL (System Demonstrations)}, 2014, pp. 55--60.

\bibitem{bird2006nltk}
S.~Bird, ``Nltk: the natural language toolkit,'' in \emph{Proceedings of the
  COLING/ACL on Interactive presentation sessions}.\hskip 1em plus 0.5em minus
  0.4em\relax Association for Computational Linguistics, 2006, pp. 69--72.

\bibitem{loria2013text}
S.~Loria, ``https://textblob.readthedocs.io/en/dev/,'' 2013.

\bibitem{liu2005opinion}
B.~Liu, M.~Hu, and J.~Cheng, ``Opinion observer: analyzing and comparing
  opinions on the web,'' in \emph{Proceedings of the 14th International
  conference on World Wide Web}.\hskip 1em plus 0.5em minus 0.4em\relax ACM,
  2005, pp. 342--351.

\bibitem{Liu2015ARS}
Q.~Liu, Z.~Gao, B.~Liu, and Y.~Zhang, ``Automated rule selection for aspect
  extraction in opinion mining,'' in \emph{Proceedings of the 24th
  International Conference on Artificial Intelligence}, ser. IJCAI'15, 2015,
  pp. 1291--1297.

\bibitem{mikolov2013distributed}
T.~Mikolov and J.~Dean, ``Distributed representations of words and phrases and
  their compositionality,'' \emph{Advances in Neural Information Processing
  systems}, 2013.

\bibitem{pesarin2012conversation}
A.~Pesarin, M.~Cristani, V.~Murino, and A.~Vinciarelli, ``Conversation analysis
  at work: detection of conflict in competitive discussions through
  semi-automatic turn-organization analysis,'' \emph{Cognitive processing},
  vol.~13, no.~2, pp. 533--540, 2012.

\bibitem{vinyals2008towards}
O.~Vinyals and G.~Friedland, ``Towards semantic analysis of conversations: A
  system for the live identification of speakers in meetings,'' in
  \emph{Semantic Computing, 2008 IEEE International Conference on}.\hskip 1em
  plus 0.5em minus 0.4em\relax IEEE, 2008, pp. 426--431.

\bibitem{jovanovic2004towards}
N.~Jovanovi{\'c} \emph{et~al.}, ``Towards automatic addressee identification in
  multi-party dialogues.''\hskip 1em plus 0.5em minus 0.4em\relax Association
  for Computational Linguistics, 2004.

\bibitem{wyatt2007privacy}
D.~Wyatt, T.~Choudhury, J.~A. Bilmes, and H.~A. Kautz, ``A privacy-sensitive
  approach to modeling multi-person conversations.'' in \emph{IJCAI}, vol.~7,
  2007, pp. 1769--1775.

\bibitem{Harper2015MovieLens}
F.~M. Harper and J.~A. Konstan, ``The movielens datasets: History and
  context,'' \emph{ACM Trans. Interact. Intell. Syst.}, vol.~5, no.~4, pp.
  19:1--19:19, Dec. 2015.

\bibitem{hofmann2001unsupervised}
T.~Hofmann, ``Unsupervised learning by probabilistic latent semantic
  analysis,'' \emph{Machine learning}, vol.~42, no. 1-2, pp. 177--196, 2001.

\bibitem{mehta2007robust}
B.~Mehta, T.~Hofmann, and W.~Nejdl, ``Robust collaborative filtering,'' in
  \emph{Proceedings of the 2007 ACM conference on Recommender systems}.\hskip
  1em plus 0.5em minus 0.4em\relax ACM, 2007, pp. 49--56.

\bibitem{schalkwyk2010your}
J.~Schalkwyk, D.~Beeferman, F.~Beaufays, B.~Byrne, C.~Chelba, M.~Cohen,
  M.~Kamvar, and B.~Strope, ``“your word is my command”: Google search by
  voice: A case study,'' in \emph{Advances in Speech Recognition}.\hskip 1em
  plus 0.5em minus 0.4em\relax Springer, 2010, pp. 61--90.

\bibitem{knijnenburg2012explain}
B.~P. Knijnenburg, M.~C. Willemsen, Z.~Gantner, H.~Soncu, and C.~Newell,
  ``Explaining the user experience of recommender systems,'' \emph{User
  Modeling and User-Adapted Interaction}, vol.~22, no. 4-5, pp. 441--504, Oct.
  2012.

\bibitem{pu2011user}
P.~Pu, L.~Chen, and R.~Hu, ``A user-centric evaluation framework for
  recommender systems,'' in \emph{Proceedings of the Fifth ACM Conference on
  Recommender Systems}, ser. RecSys '11.\hskip 1em plus 0.5em minus 0.4em\relax
  New York, NY, USA: ACM, 2011, pp. 157--164.

\end{thebibliography}

\end{document}